\newif\iflandscape
\newif\ifportrait
\typein[\lorp]%
{Typein "l" (for landscape, twocolumn) or "p" (for portrait, onecolumn)}
\if l\lorp \landscapetrue \else \portraittrue \fi
\newlength{\extralineskip}
\ifportrait
  \documentstyle[12pt]{article}
\fi
\iflandscape
  \documentstyle[twocolumn]{article}
\fi
\def\tr#1{{\rm tr}\kern-3pt\left[#1\right]}
\def\stdi{\strut\displaystyle}
\def\ooo#1#2{{\stdi#1\over\stdi#2}}
\def\D{{\cal D}}
\begin{document}
\begin{titlepage}
\begin{flushright}
          \begin{minipage}[t]{12em}
          \large UAB--FT--306   \\
                 hep-th/9304090 \\
                 April 1993
          \end{minipage}
\end{flushright}

\vspace{\fill}

\vspace{\fill}

\begin{center}
\baselineskip=2.5em
{\huge RENORMALIZATION GROUP APPROACH TO MATRIX MODELS\\
       IN TWO-DIMENSIONAL QUANTUM GRAVITY$^*$}
\end{center}

\vspace{\fill}

\begin{center}
{\sc Carles AYALA}$^{\dag}$\\
     Grup de F\'\i sica Te\`orica and Institut de F\'\i sica d'Altes Energies\\
     Universitat Aut\`onoma de Barcelona\\
     08193 Bellaterra, Barcelona, Spain
\end{center}

\vspace{\fill}

\begin{center}
\large ABSTRACT
\end{center}
\begin{center}
\begin{minipage}[t]{36em}
We explore the implications of recent work by Br\'ezin and Zinn-Justin,
applying the renormalization group techniques from critical phenomena to the
scaling limit of matrix models in two-dimensional quantum gravity.
They endeavor to get the lowest order fixed points of the theory
giving insight upon the critical points of the theory.
We show that at leading order the perturbative result is equal to the
saddle-point approximation result.
We calculate the next-to-leading order in the perturbative expansion
exploring the goodness of the approach.
\end{minipage}
\end{center}

\vspace{\fill}

{\noindent\makebox[10cm]{\hrulefill}\\
\footnotesize
\makebox[1cm][r]{$^{\dag}$} Research supported by a CIRIT-DGU postdoctoral
                            fellowship BPOST91-13 (also BPOST92-18)\\
\makebox[1cm][r]{\        } E-mail address: {\tt ayala$@$ifae.es} or
                                            {\tt ifte1$@$cc.uab.es}\\[0.2cm]
\makebox[1cm][r]{     $^*$} Work partially supported by Research Project CICYT.
}

\end{titlepage}

\clearpage

\def\tableone{
\begin{table}
\centering
\begin{tabular}{|l|l|}\hline
$g^\ast_{4},\ g^*_{6},\ g^*_{22}$&
Eigenvalues of $\ooo{\partial\beta_i}{\partial g_j}(g^\ast)$     \\ \hline
\hline
$-0.1011768497154676962372589620950$,  & $1.217$                 \\
$-0.0050734359765849695624121167177$,  & $-0.894 + 0.0389 i$     \\
$-0.0063999132724733397458433406611$   & $-0.894 - 0.0389 i$     \\ \hline
\hline
$1.24418126545378004447923811987691$,  & $4.402+6.361i$          \\
$-0.7769430030324647493548664168342$,  & $4.402-6.361i$          \\
$-2.5952930639996386407825041596420$   & $-5.3874$               \\ \hline
\hline
$0.52437110813928673984961366257774$,  & $0.217+ 2.330i$         \\
$-0.1640898595547203137714211279754$,  & $0.217- 2.330i$         \\
$-1.3720478532685124859943450768933$   & $2.508$                 \\ \hline
\hline
$-3.6625519429210568815087489975509$,  & $35.714$                \\
$3.42697568626088468656019000784756$,  & $11.530+25.953i$        \\
$ -1.061522619035009390201712691217$   & $11.530-25.953i$        \\ \hline
\hline
$2.05544017071028753922765070814953$,  & $15.171 + 13.9986i$     \\
$-0.4847069709465395570725370917872$,  & $15.171 - 13.9986i$     \\
$0.45720234798000021436376439464735$   & $11.5665955$            \\ \hline
\hline
$0$, $0$, $0$                          & $-2$, $-2$, $-1$        \\ \hline
\end{tabular}
\caption[\tableonetext]{}
\vspace{1ex}
\begin{minipage}{20em}
\tableonetext
\end{minipage}
\label{TableOne}
\end{table}
}
\def\tableonetext{
All real zeros for the $\beta_4$, $\beta_6$,
$\beta_{22}$ and the eigenvalues of the  matrix
$\partial\beta_i/\partial g_j$.  The first row
is the interesting solution. The solution at the zero is included as a
curiosity, and the rest are shown for completeness.
}

\addtolength{\baselineskip}{\extralineskip}
\vbox{\vspace{6em}}

\begin{center}
\section{Introduction}
\end{center}

Recently \cite{BZ}, Br\'ezin and Zinn-Justin proposed to apply the apparatus
of renormalization group and critical phenomena to the new development of two
dimensional quantum gravity coming out from its transcription to a matrix
model \cite{David,Jerusalem}. In fact they achieve to outline,
qualitatively and semi-quantitatively, the critical
points and critical exponents of the theory.
That approach have raised interest and its method and spirit has been applied
elsewhere \cite{Alfaro,Periwal,Gao,Japs}.

In two dimensional quantum gravity a cumbersome situation is faced. The
solvability of the theory is dramatically dependent on the range of values
taken by the central charge $c$ of the matter field. After \cite{BKDSGM},
we learn to solve exactly the problem for $c\leq 1$. Unfortunately, for $c>1$,
despite the possibility to construct matrix models for those
cases \cite{BH}, we cannot
solve the problem. Moreover, we do not know if there is a continuum theory.

In \cite{BZ}, a procedure is build to attempt to extract information from the
$c\leq1$ cases, which are exactly solvable, that could be exported to the $c>1$
ones. They check the procedure in the $c=0$ case at the lowest
order using two calculation methods: perturbative expansion and saddle point
approximation (SPA). The results they found for the fixed points obtained
are calculation method dependent. We will show that the perturbative one
gives the same
result as the SPA. Gladly, we can read in \cite{Alfaro} that this result
appear again in the $c=1$ case \cite{BKZ}.
The fundamental point in the exact solution of \cite{BKDSGM} is the
existence of a ``double scaling limit''. That is, the matrix coupling constant
$g$ and the size $N$ of the matrix considered follow a scaling limit. It
consists in making $g\rightarrow0$ and   $N\rightarrow\infty$ in a way that
$(g-g_c)N^{2/\gamma_1}$ is finite. The procedure presented in \cite{BZ} is
that the evolution of $N$, $N\mapsto N+\delta N$ must be compensated by a
change of $g$, $g\mapsto g+\delta g$, having the same continuum physics. This
unfortunately will force us to enlarge the space of coupling constants, as in
the Wilson's scheme \cite{Wilson}. So, in its way to the scaling
limit the string partition function must obey a Callan-Symanzik like equation
\begin{equation}
\left[N{\partial\over\partial N}
 - \beta( g){\partial\over\partial g}
+ \gamma(g)\right] Z(N,g) = r(g)\ .
\label{scaling}
\end{equation}
with a fixed point $g^*$ given by $\beta(g^*)=0$ and $\beta^\prime(g^*)>0$.

As we learned from \cite{plenty} the scaling laws for $c<1$ forces the
singular part of the string partition function to follow
\begin{equation}
Z = (g_c-g)^{ 2-\gamma_ 0}f \left((g_c-g) \ N^{2/\gamma_ 1}\right) \ ,
\label{Zpara}
\end{equation}
where $g_c$ is the critical point and
\begin{displaymath}
\gamma_ 1 = 2 - \gamma_ 0 = {25-c+ \sqrt{( 1-c)(25-c)}\over12}\ \ .
\end{displaymath}
The string susceptibility at genus $h$ is
\begin{displaymath}
\gamma_h = \gamma_ 0 + h \gamma_1\ \ .
\end{displaymath}

Introducing the scaling law (\ref{Zpara}) in eq. (\ref{scaling}), one obtains
the following constraints
\begin{displaymath}
\gamma_1 = {2\over\beta^\prime(g^*)},\ \ \ \
\gamma_0 = 2 - {\gamma(g^*)\over\beta^\prime(g^*)} \ .
\label{gammas}
\end{displaymath}

In the exact solution $g_c=-1/12$ and $\gamma_1=5/2$.

In \cite{BZ},
the building of the flow equations for the migration from $(N+1)\times(N+1)$
matrix $\phi_{N+1}$ to  the $N\times N$ matrix  $\phi_{N}$ is done by
integrating out one row and one column in $\phi_{N+1}$ . They see that the
matrix partition function $\zeta_N(g)$ follow the equation
\begin{displaymath}
\zeta_{N+1}(g)=\left\{\lambda(g)\right\}^{N^2}\zeta_N(g^\prime)\ ,
\end{displaymath}
and the string partition function $Z(N,g)$ can be written as
\begin{displaymath}
Z(N,g) = {1\over N^2}\log\zeta(N,g)
\end{displaymath}
using (\ref{scaling}) they find
\begin{equation}
\begin{array}{rl}
g^\prime   & = g + \ooo{1}{N}\beta(g) + O\left(\ooo{1}{N^2}\right),\\[1ex]
\lambda(g) & = 1 + \ooo{1}{N}r(g) +     O\left(\ooo{1}{N^2}\right).
\end{array}
\label{functions}
\end{equation}

The organization of the paper is as follows. In Sect. 2, we work out the lowest
order. In Sect. 3, we present some technicalities needed to work out higher
orders. In Sect. 4, the results for the $O(g^3)$ are presented and analyzed.
Finally, in Sect. 5 we present our conclusions.

\begin{center}
\section{The lowest order}
\end{center}
As presented in \cite{BZ}, the matrix $\phi_{N+1}$ is written in terms of an
$N\times N$ submatrix $ \phi_N$,
a complex $ N $-component vector $ v_a,$ and a number $\alpha$:
\begin{equation}
\phi_{N+1} = \left( \matrix{\stdi \phi_ N  &\stdi v_a   \cr
                            \stdi v^\ast_a &\stdi \alpha\cr} \right)
\label{para}
\end{equation}
and the renormalization flow is obtained by integrating over the $v_a$
and $v^\ast_a$. All terms related with $\alpha$ are of relative order
$1/N$\cite{BZ}, therefore we can set $\alpha=0$.

Our original action is
\begin{equation}
S_{N+1}(\Phi_N,g,v) =(N+1)\tr{\ooo{1}{2}\Phi_{N+1}^2 + \ooo{g}{4}\Phi_{N+1}^4}
\ .
\label{OriginalAction}
\end{equation}

After all the expansion is done we will get an effective action as
\begin{equation}
\begin{array}{rl}
S_{N+1}(\Phi_N,g) =&
(N+1)\tr{{\stdi1\over\stdi2}\Phi_N^2 + {\stdi g\over\stdi4}\Phi_N^4}
+ C_2\tr{\Phi_N^2}
+ C_4\tr{\Phi_N^4}\\[1ex]
=&\left({\stdi N+1\over\stdi2} + C_2\right)\tr{\Phi_N^2}
+\left({\stdi N+1\over\stdi4}g + C_4\right)\tr{\Phi_N^4} \ .
\end{array}
\label{twoaction}
\end{equation}
Our renormalization prescription is that the $\tr{{\Phi^\prime}_N^2}$ term must
have a $N/2$ coefficient in such a way that
\begin{displaymath}
S_{N+1}(\Phi_N^\prime,g^\prime) =
N\tr{{1\over2}{\Phi^\prime}_N^2 + {g^\prime\over4}{\Phi^\prime}_N^4}\ ,
\end{displaymath}
using the definition
\begin{equation}
\Phi_N=\rho\Phi^\prime_N\ \ .
\label{rhoori}
\end{equation}

Eventually, one can write
\begin{equation}
\begin{array}{rl}
\rho =&          \left(1 + {\stdi 1+2C_2\over\stdi N}\right)^{-1/2}\ ,\\[1ex]
g^\prime=& \rho^4\left(g + {\stdi g+4C_4\over\stdi N}\right)\ .
\end{array}
\label{rhog}
\end{equation}
Thus, expanding both quantities till $O(1/N)$
\begin{displaymath}
\begin{array}{rl}
   \rho =& 1 - {\stdi1+2C_2\over\stdi 2N} +
           O\left({\stdi1\over\stdi N^2}\right)\ ,\\[1ex]
g^\prime=& g + {\stdi-g+4C_4-4C_2g\over\stdi N} +
           O\left({\stdi1\over\stdi N^2}\right)\ .
\end{array}
\end{displaymath}
{}From that last formulae we can see that the use of $C_2$ at $O(g)$ forces us
to
include $C_4$ at order $O(g^2)$ (missed in \cite{BZ}). Calculating those
quantities at the lowest order
\begin{displaymath}
C_2 = g -g^2 +\cdots \ \ \ \ C_4 = - {1\over 2}g^2 + \cdots\ .
\end{displaymath}
And now
\begin{displaymath}
g^\prime= g + {-g-6g^2\over N}\ ,
\end{displaymath}
as the result of the saddle point approximation (SPA) \cite{BZ}. Thus
\begin{equation}
\beta(g) = -g-6g^2\ .
\label{betag}
\end{equation}
And the resulting fixed point is $g^*=-1/6$
(that gives $\beta^\prime(g^*)>0$).
This is the first result of our paper. We understand it realizing
that both results (``perturbative'' and SPA) are
obtained till the same
order of expansion in $g$ in the same regime of
the theory (they had to be equal).

\begin{center}
\section{$\bf O(g^k)$: some useful technology}
\end{center}

The reasonable behavior of the first order naive calculation gave us a certain
hope that the computation of the next orders will give some meaningful results.
As stated in \cite{BZ} and mentioned earlier in Sect. 1, a naive
expansion in higher orders of $g$ will make
appear new terms that must be added to the action. The coefficients (coupling
constants) of those terms must be taken into account in the renormalization
procedure. The goal of the present section is to develop a explicit
construction to calculate any given order.
The replacement for (\ref{scaling}) will be
\begin{equation}
\left[N{\partial\over\partial N}
 - \sum_i\beta_i( g_j){\partial\over\partial g_i}
+ \gamma(g)\right] Z(N,g) = r(g)
\end{equation}
with a fixed point $g^*_i$ given by $\beta_k(g^*_i)=0$ and that all the
eigenvalues of the matrix formed as
$\partial\beta_l/\partial g_m(g^*_i)$
are positive.

If we consider $O(g^3)$, the new terms are $\tr{\phi^6}$ and
$\tr{\phi^2}^2$.
At $O(g^4)$ the new terms are $\tr{\phi^8_N}$ and
$\tr{\phi^2_N}\tr{\phi^4_N}.$%
\footnote{
We could see the same in the SPA by further
expanding  the formula for the $\sigma$ (Lagrange multiplier field that fulfil
the SPA) in \cite{BZ}
\begin{displaymath}
    \sigma={1\over N}\tr{1\over 1 + g\sigma + g\phi^2_N}
\end{displaymath}
giving
\begin{displaymath}
    \begin{array}{l}
     \sigma=1-g\left(1+t_2\right)
             +g^2\left(2+3t_2+t_4\right)
             -g^3\left(5+10t_2+4t_4+t_6+2(t_2)^2\right)\\[0.4ex]
     \hspace{6em}
     +g^4\left(14+35t_2+15t_4+5t_6+15(t_2)^2+t_8+5t_2t_4\right)+\cdots
    \end{array}
\end{displaymath}
where we defined
\begin{displaymath}
t_n=\ooo{1}{N}\tr{\phi^n_N}.
\end{displaymath}
}
Just to fix ideas, at $O(g^3)$
\begin{displaymath}
\begin{array}{rl}
S_{N+1}(\Phi_N,g) = \left( \ooo{N+1}{2}+g-g^2
                          +\ooo{2N+3}{N+1}g^3\right)
                          \tr{\phi^2_N}&\\[0.4ex]
                  + \left( \ooo{N+1}{4}g
                          -\ooo{1}{2}g^2
                          +\ooo{2N+3}{2(N+1)}g^3\right)
                          \tr{\phi^4_N}&\\[1ex]
                  +       \ooo{1}{3}g^3      \tr{\phi^6_N}
                  +       \ooo{1}{2(N+1)}g^3 \tr{\phi^2_N}^2\ ,
\end{array}
\end{displaymath}
where we can see that at $O(g^2)$ no new terms must be included to the action.

As the order of expansion increases we need to know the behaviour of higher
order terms under eq. (\ref{para}). Thus, let us  assume an original action
$S_N^{(k)}(\Phi_N,g)$ needed to workout the $O(g^k)$. The cases
that will be shown in the present section are two orders more apart
from the lowest one
\begin{equation}
S_N^{(2)}(\Phi_N,g) =
N\left(
\ooo{1}{2}\tr{\phi_N^2}
+\ooo{g_{4}}{4}\tr{\phi_N^{4}}
\right)
\label{BigActionZero}
\end{equation}
presented and solved in the previous section.
Those two new actions are presented as a helpful way to understand the
constructive quality of the procedure to be shown bellow. First
\begin{equation}
S_N^{(3)}(\Phi_N,g) =
N\left(
\ooo{1}{2}\tr{\phi_N^2}
+\ooo{g_{4}}{4}\tr{\phi_N^{4}}
+\ooo{g_{6}}{6}\tr{\phi_N^{6}}
\right)
+\ooo{g_{22}}{4}\tr{\phi_N^2}^2
\label{BigActionOne}
\end{equation}
which involves three coupling constants: $g_4$, $g_6$ and $g_{22}$. Then, to
analyze the $\beta_i$ we
will have to find the roots of mixed polynomials with maximum power three in
those coupling constants, that is to say terms as  $g_4^k g_6^l g_{22}^m$ with
$k+l+m\leq3$. The next order is a slightly more messy, including five coupling
constants as can be seen in
\begin{equation}
\begin{array}{rl}
S_N^{(4)}(\Phi_N,g) =%
&N\left(
\ooo{1}{2}\tr{\phi_N^2}
+\ooo{g_{4}}{4}\tr{\phi_N^{4}}
+\ooo{g_{6}}{6}\tr{\phi_N^{6}}
+\ooo{g_{8}}{8}\tr{\phi_N^{8}}\right)\\[2ex]
&
\hspace{0.5cm}+\ooo{g_{22}}{4}\tr{\phi_N^2}^2
+\ooo{g_{24}}{8}\tr{\phi_N^2}\tr{\phi_N^4},
\end{array}
\label{BigActionTwo}
\end{equation}
whose complexity is a consequence of the renormalization procedure.

As in the last section we start from the action $S_{N+1}^{(k)}(\Phi_{N+1},g)$.
We evolve them to $S_N^{\prime(k)}(\Phi^{\prime}_N,g^{\prime})$
action due to use of two steps.

The first step consists in how the $\tr{\phi_{N+1}^n}$
terms change with the parameterization written in eq. (\ref{para}), so far
\begin{equation}
\tr{\phi_{N+1}^2} = \tr{\phi_N^2} + 2v^\ast v\label{square}
\end{equation}
\begin{equation}
\begin{array}{rl}
\tr{\phi_{N+1}^4} = &
\tr{\phi_N^4} + 4v^\ast\phi_N^2 v\\[1ex]
         &+ 2(v^\ast v)^2
\end{array}
\end{equation}
\begin{equation}
\begin{array}{rl}
\tr{\phi_{N+1}^6} = &
\tr{\phi_N^6} + 6v^\ast\phi_N^4 v\\[1ex]
         &+ 6(v^\ast v)(v^\ast\phi_N^2 v)
          + 3(v^\ast\phi_N v)^2\\[1ex]
         &+ 2(v^\ast v)^3
\end{array}
\end{equation}
\begin{equation}
\begin{array}{rl}
\tr{\phi_{N+1}^8} = &
\tr{\phi_N^8} + 8v^\ast\phi_N^6 v\\[1ex]
         &+ 8(v^\ast v)(v^\ast\phi_N^4 v)
          + 8(v^\ast\phi_N v)(v^\ast\phi_N^3 v)\\[1ex]
         &+ 4(v^\ast\phi_N^2 v)^2
          + 8(v^\ast v)^2(v^\ast\phi_N^2 v)\\[1ex]
         &+ 8(v^\ast v)(v^\ast\phi_N v)^2
          + 2(v^\ast v)^4\label{eight}
\end{array}
\end{equation}
and for the composite terms
\begin{equation}
\begin{array}{rl}
\tr{\phi_{N+1}^2}^2 = & \tr{\phi_N^2}^2\\[1ex]
&+4(v^\ast v){\rm tr}\left[\phi_N^2\right]+ 4(v^\ast v)^2\ ,\\[2ex]
\end{array}
\label{comptwotwo}
\end{equation}
\begin{equation}
\begin{array}{rl}
\tr{\phi_{N+1}^2}\tr{\phi_{N+1}^4} = &
\tr{\phi_N^4}\tr{\phi_N^2}\\[1ex]
 &+2(v^\ast v)\tr{\phi_N^4}
  +4\tr{\phi_N^2}(v^\ast\phi_N^2 v)\\[1ex]
 &+2(v^\ast v)^2\tr{\phi_N^2}
  +8(v^\ast\phi_N^2 v)(v^\ast v)
  +4(v^\ast v)^3\ .
\end{array}
\label{comptwofour}
\end{equation}
In a general case, and looking at the structure of eqs.
(\ref{square})--(\ref{eight}), we can write
\begin{equation}
\begin{array}{rl}
S_{N+1}^{(k)}(\Phi_N,g) = &\ooo{N+1}{N}S_N(\Phi_N,g)\nonumber\\[2ex]
&+(N+1)v^\ast\left\{1+g_4\phi^2+g_6\phi^4+g_8\phi^6+\cdots\right\}v
  \label{struct}\\[2ex]
& \hspace{6em}+\Delta S(\Phi_N,g,v,v^\ast)
\nonumber
\end{array} \label{NewAction}
\end{equation}

The second step of the migration of
the $S_{N+1}^{(k)}(\Phi_{N+1},g)$ actions to the
$S_N^{\prime(k)}(\Phi^{\prime}_N,g^{\prime})$
is the result of the integration
\begin{displaymath}
\exp[-S_{N}^\prime(\Phi^\prime_N,g^\prime)]=
\left(\lambda(g)\right)^{N^2}\int dv\,dv^\ast\,
\exp[-S_{N+1}(\Phi_N,g,v,v^\ast)]
\end{displaymath}

In our case, far from the exact result of \cite{BKDSGM}, we are interested
in a perturbative expansion on the $g_4$, $g_6$,... As usual, we can
pretend to construct a generating functional of the problem. A careful look at
eq. (\ref{struct}) reveals that a simple and clear possibility is
\begin{equation}
\begin{array}{rl}
I(x,\phi_N) \equiv&\stdi\int d v^\ast d v
\exp\left[-(N+1)
v^\ast\left\{\sum_{n\geq0} x_n\phi_N^n\right\}v
\right]
\nonumber\\[4ex]
\propto&
\exp\left[
-\tr{\log\left\{\stdi\sum_{n\geq0} x_n\phi_N^n\right\}}
\right]\ .
\label{I}
\end{array}
\end{equation}

To generate $S_{N+1}(\Phi_N,g)$ from $I(x,\phi_N)$ we will show the need to
construct a multiplicative differential operator $\D$, as follows
\begin{equation}
\begin{array}{rl}
&\exp\left[-S_{N+1}(\Phi_N,g)\right] \equiv\\[3ex]
&\ \ \ \
\exp\left[-\ooo{(N+1)}{N}S_N(\Phi_N,g)\right]
{\displaystyle\lim
\atop
\begin{array}{l}
      \strut\scriptstyle x_{0}    \to 1       \\
      \strut\scriptstyle x_{2n+1} \to 0       \\
      \strut\scriptstyle x_{2n}   \to g_{2n+2}
\end{array}
}
\exp\left[{{\cal D}}\right]I(x,\phi_N)\ ,
\end{array}
\label{gene}
\end{equation}
The role played by the operator ${\cal D}$ is to represent all the extra terms
in eqs. (\ref{square})--(\ref{comptwofour}) written as
$\Delta S(\Phi_N,g,v,v^\ast)$ in eq. (\ref{NewAction}).
The final result of the limit in eq. (\ref{gene}) will be the addition of new
terms to the action in the following way
\begin{equation}
\Delta S = \sum_{n\geq0}C_{2n}\tr{\phi_N^{2n}} +
           \sum_{n,m}C_{2n,2m}\tr{\phi_N^{2n}}\tr{\phi_N^{2m}} + \cdots
\label{incact}
\end{equation}
as was presented at $O(g^2)$ in eq. (\ref{twoaction}). Then
\begin{equation}
\begin{array}{rl}
S_{N+1}(\phi^\prime_N,g^\prime) =&
\left\{\ooo{N+1}{2} + C_2\right\}\tr{\phi_N^2} + \cdots\\[2ex]
&+\left\{(N+1) \ooo{g_{2n}}{2n} + C_{2n}\right\}
                             \tr{\phi_N^{2n}} + \cdots\\[2ex]
&+\left\{\ooo{g_{2n,2m}}{4nm} + C_{2n,2m}\right\}
                             \tr{\phi_N^{2n}}\tr{\phi_N^{2m}} + \cdots
\end{array}
\end{equation}
Following the same prescription as in the lowest order case,
therefore using eq. (\ref{rhoori}) and the same from for the $\rho$
as in eq. (\ref{rhog}) but
\begin{equation}
\begin{array}{l}
g^\prime_{2n}=\left(g_{2n} + \ooo{ g_{2n}+2nC_{2n}}{N}\right)\rho^{2n}\\[2ex]
g^\prime_{2n,2m}=\left(g_{2n,2m} + 4nmC_{2n,2m}\right)\rho^{4nm}\\[2ex]
\end{array}
\label{gp}
\end{equation}
where $C_{2n}$ is to be collected at $O(1/N^0)\equiv O(1)$ and
$C_{2n,2m}$ at $O(1/N)$.
Now the definition of the $\beta_T$ translate to
\begin{equation}
\beta_g = N\left(g^\prime - g\right) \ .
\end{equation}

The core of the initiative to construct the generating function  in
eq. (\ref{I}),  followed two requirements. First, the calculability of the
generating function itself. As written in the same
equation it can be constructively expanded till any given order.
Last, the
possibility to construct operators that will reproduce the terms of the
original
action, outlined in the last line of eq.
(\ref{NewAction}) as $\Delta S(\Phi_N,g,v,v^\ast)$.
That is
\begin{equation}
\begin{array}{rl}
&\exp\left[C_i\ooo{\partial}{\partial x_i}\right]
I(x,\phi_N) =  \\[2ex]
&\hspace{2em}\stdi\int d v^\ast d v
\exp\left[
-(N+1)\left(C_i  v^\ast\phi_N^i v + v^\ast\sum_{n\geq0} x_n\phi_N^nv\right)
\right]
\end{array}
\label{oneder}
\end{equation}
\begin{equation}
\begin{array}{rl}
&\exp\left[-\ooo{C_{ij}}{N+1}\ooo{\partial^2}{\partial x_i\partial x_j}\right]
I(x,\phi_N) =  \\[2ex]
&\hspace{2em}\stdi\int d v^\ast d v
\exp\left[
-(N+1)\left( C_{ij}  (v^\ast\phi_N^i v) (v^\ast\phi_N^j v)
            +v^\ast\sum_{n\geq0} x_n\phi_N^nv
      \right)
\right]
\end{array}
\label{twoder}
\end{equation}

No sum is implicitly understood in those formul\ae, although the repeated
indexes could also be taken as an implicit sum. Eqs. (\ref{oneder}) and
(\ref{twoder}) can be resumed in the following substitution
\begin{equation}
(v^\ast\phi_N^m v) \mapsto -\ooo{1}{N+1}\ooo{\partial}{\partial x_m}
\label{subs}
\end{equation}

Using eq. (\ref{subs}), we can construct in any order
in the expansion on the $g$'s the operator $\D$ in eq. (\ref{gene}).

In the same notation as (\ref{BigActionOne}) and (\ref{BigActionTwo}) we
propose $\D$ and writing explicitly till the $O(g^4)$ order
\begin{equation}
\begin{array}{rl}
&\D^{(2)}=\ooo{g_4}{4}\D_4\\[1ex]
&\D^{(3)}=\ooo{g_4}{4}\D_4 + \ooo{g_6}{6}\D_6 +\ooo{g_{22}}{4}\D_{22}\\[1ex]
&\D^{(4)}=\ooo{g_4}{4}\D_4 + \ooo{g_6}{6}\D_6 +\ooo{g_{22}}{4}\D_{22}
          + \ooo{g_8}{8}\D_8
          + \ooo{g_{24}}{8}\D_{24}
\end{array}
\end{equation}
where all the $\D_A$ represent the contribution of terms
but the two first ones of each of the RHS of eqs.
(\ref{square})--(\ref{eight}), or the contribution of all terms but the first
in the RHS in eqs.  (\ref{comptwotwo}) and (\ref{comptwofour}).  Namely, for
the lowest order the only operator is
\begin{displaymath}
\begin{array}{rl}
\D_4=& 2\epsilon  \ooo{\partial^2}{\partial x_0^2}\\[2ex]
\end{array}
\end{displaymath}
where we defined
\begin{displaymath}
\epsilon=-\ooo{1}{N+1}\ ,
\end{displaymath}
already in the next-to-leading order, two far more complicated operators
are needed
\begin{displaymath}
\begin{array}{rl}
\D_6=& 6\epsilon  \ooo{\partial^2}{\partial x_0\partial x_2}
      +3\epsilon  \ooo{\partial^2}{\partial x_1^2}
      +2\epsilon^2\ooo{\partial^3}{\partial x_0^3}\\[2ex]
\end{array}
\end{displaymath}
\begin{displaymath}
\begin{array}{rl}
\D_{22}=&-4\epsilon  \tr{\phi_N^2}\ooo{\partial}{\partial x_0}
         -4\epsilon^2\ooo{\partial^2}{\partial x_0^2}\\[2ex]
\end{array}
\end{displaymath}
as we go on that complexity can only increase, as we can see for the $O(g^4)$,
as follows
\begin{displaymath}
\begin{array}{rl}
\D_8=& 8\epsilon\ooo{\partial^2}{\partial x_0\partial x_4}
      +8\epsilon\ooo{\partial^2}{\partial x_1\partial x_3}
      +4\epsilon\ooo{\partial^2}{\partial x_2^2}\\[1ex]
     &+8\epsilon^2\ooo{\partial^3}{\partial x_0^2\partial x_2}
      +8\epsilon^2\ooo{\partial^3}{\partial x_0\partial x_1^2}
      +2\epsilon^3\ooo{\partial^4}{\partial x_0^4}\\[2ex]
\end{array}
\end{displaymath}
\begin{displaymath}
\begin{array}{rl}
\D_{24}=&-2\epsilon  \tr{\phi_N^4}\ooo{\partial}{\partial x_0}
         -4\epsilon  \tr{\phi_N^2}\ooo{\partial}{\partial x_2}
         -2\epsilon^2\tr{\phi_N^2}\ooo{\partial^2}{\partial x_0^2}\\[1ex]
        &-8\epsilon^2\ooo{\partial^2}{\partial x_0\partial x_2}
         -4\epsilon^3\ooo{\partial^3}{\partial x_0^3}
\end{array}
\end{displaymath}
At that point, we made use of two of the most glorious achievements in computer
algebra\cite{REDUCE,Mathematica} to analyze the problem.
We can transcript the problem and get the
results avoiding the overwhelming and tedious algebra.

In that section we have fulfilled our quest for a perturbative expansion in
terms of the computation of eq. (\ref{gene}). That means to write the
appropriate $\D$ and expand both the
exponential of $\D$ and the $I(x,\phi)$ for the order required.

\begin{center}
\section{$O(g^3)$: results}
\end{center}

Unleashing all the developed technology in the last section we can find the
following results for the $\beta$ functions of the next-to-leading order.
The increment to the action in eq. (\ref{NewAction}) is
\def\GA{g_{4}}  %
\def\GB{g_{6}}  %
\def\GC{g_{22}} %
\begin{equation}
\begin{array}{l}
\Delta S_{N+1}(\Phi_N,g,v,v^\ast) =
         (N+1)\GA\ooo{1}{2}(v^\ast v)^2 \nonumber\\[2ex]
\makebox[3em][r]{$+$}(N+1)\GB\left[(v^\ast\phi_N^4 v) +
                   (v^\ast\phi_N^2 v)(v^\ast v) +
                   \ooo{1}{2}(v^\ast\phi_N v)^2  +
                   \ooo{1}{3}(v^\ast v)^3\right] \nonumber\\[2ex]
\makebox[3em][r]{$+$}\GC\left[\ooo{1}{2}(v^\ast v)\tr{\phi_N^2} +
                           (v^\ast v)^2\right]\ . \nonumber\\[4ex]
\end{array}
\end{equation}
The resulting formul\ae\ are
\begin{equation}
\begin{array}{rl}
\beta_{\GA}=&-\GA-6 \GA^{2}+4 \GB-6 \GB^{2}-12 \GA \GB-4 \GA \GC    \\[1ex]
            &+8 \GA^{3}+36 \GA^{2} \GB+4 \GA^{2} \GC+48 \GA \GB^{2} \\[1ex]
            &+4 \GA \GB \GC+20 \GB^{3}\ ,
\end{array}
\label{BetaG4}
\end{equation}
\begin{equation}
\begin{array}{rl}
\beta_{\GB}=&2 \left(-\GB-6 \GB^{2}-3 \GB \GC-6 \GA \GB \right.   \\[1ex]
            &+\GA^{3}+12 \GA^{2} \GB+27 \GA \GB^{2}               \\[1ex]
            &\left.+3 \GA \GB \GC+16 \GB^{3}+3 \GB^{2} \GC\right) \\[1ex]
\end{array}
\label{BetaG6}
\end{equation}
and
\begin{equation}
\begin{array}{rl}
\beta_{\GC}=
&2 \left(-\GC-3\GB^{2}-3\GC^{2}-2\GA\GB-4\GA\GC-6\GB\GC\right.\\[1ex]
&\GA^{3}+12 \GA^{2} \GB+8 \GA^{2} \GC+30 \GA \GB^{2}+30 \GA \GB \GC\\[1ex]
&\left.+5 \GA \GC^{2}+20 \GB^{3}+24 \GB^{2} \GC+6 \GB \GC^{2}\right)
\end{array}
\label{BetaG22}
\end{equation}
Also, remembering eq. (\ref{functions})
\begin{displaymath}
\begin{array}{l}
\makebox[6em][r]{$r(g)=$}
\ooo{1}{6}\left(-2 \GB-3 \GA+3 \GA^{2}+3 \GB^{2}+6 \GA \GB\right. \\[1ex]
\makebox[6em][r]{\ }
\left.-5 \GA^{3}-18 \GA^{2} \GB-21 \GA \GB^{2}-8 \GB^{3}\right)   \\[1ex]
\end{array}
\end{displaymath}
That set of equations is our second result.
The zeros for all those functions can be found in Table \ref{TableOne}.
The function in eq. (\ref{BetaG4}) reduces to eq. (\ref{betag})  if we
go down to $O(g^2)$ and one set $\GB = \GC = 0$.

The only point that could need a revision is the statement that we set
$\alpha=0$ in eq. (\ref{para}). The argument that the terms in  $\alpha$ only
give contributions at order $1/N$ when calculating $O(g^2)$ is clear. As we can
read in eq. (\ref{gp}), any contributions higher than $O(1/N)$ are irrelevant
for the computation of the several $\beta$ functions.
But maybe at $O(g^3)$ those  $\alpha$ terms can give rise to new terms that
give contribution to the $\beta$ functions. Naively, we can find several terms
that could possibly give that sort of contributions. After careful and rather
tough computations that naive guess is proven wrong.
In spite of the naive first look, those terms do not give any contribution to
the  $\beta$ functions, being eqs. (\ref{BetaG4})-(\ref{BetaG22}) the final
result at the present order.

\iflandscape
   \vspace{\fill}
   \newpage
   \vspace{\fill}
   \tableone
   \vspace{\fill}
\fi

\begin{center}
\section{Discussions and conclusions}
\end{center}

As a start, we can qualify the goodness of the method as
excellent at the lowest order, where we showed that both methods used
(perturbative expansion and saddle point approximation)
give the same result for the $\beta(g)$ function for our $c=0$ case.

At the following order the results presented in Table \ref{TableOne},
need some digestion.
Reading from \cite{JZJ}, we analyze the result by thinking that the action
given in (\ref{BigActionOne}) is completely unrestricted.
Although we are focused in the problem represented by
eq. (\ref{BigActionZero}) we computed the case represented by
eq. (\ref{BigActionOne}).
So, all coupling constants are treated as the same footing.
Then the first set of roots in Table \ref{TableOne} is the
interesting one, due to the fact that corresponds to a situation where one of
the eigen-operators is relevant (positive real part of the eigenvalue) and the
rest are irrelevant (negative real part of the eigenvalue).
Thus, one coupling prevails and the other two go to zero.
That situation match our problem where one coupling constant is expected
to be relevant, as can be seen in eq. (\ref{BigActionZero}).

Again, we can qualify the method as good. The $g^\ast_4=-0.101...$ is a big
step from the first order result, $g^\ast_4=-1/6=-0.166...$,
to the exact result $g^\ast=-1/12=-0.086...$.
Further orders are expected to give improvements of that behavior and they can
be systematically calculated,
as demonstrated in Sect. 3. Obviously they will involve
more couplings and the $\beta$-functions will be messier and much bigger, but
no new conceptual difficulty will arise.

We acknowledge a enlightening discussion with Mari\`a Baig.

\vspace{4em}

\listoftables

\ifportrait

  \newpage
  \tableone

\fi

\end{document}